# ESPEI for efficient thermodynamic database development, modification, and uncertainty quantification: application to Cu-Mg

Brandon Bocklund[1,*], Richard Otis[2], Aleksei Egorov[3], Abdulmonem Obaied[3], Irina Roslyakova[3], Zi-Kui Liu[1]

[1] Department of Materials Science & Engineering, Pennsylvania State University, University Park, PA, 16802, USA

[2] Engineering and Science Directorate, Jet Propulsion Laboratory, California Institute of Technology, Pasadena, CA 91109, USA

[3] ICAMS, Ruhr-University Bochum, Universitätstr. 150, 44801, Bochum, Germany

**Abstract**

The software package ESPEI has been developed for efficient evaluation of thermodynamic model parameters within the CALPHAD method. ESPEI uses a linear fitting strategy to parameterize Gibbs energy functions of single phases based on their thermochemical data and refine the model parameters using phase equilibrium data through Bayesian optimization within a Markov Chain Monte Carlo machine learning approach. In this paper, the methodologies employed in ESPEI are discussed in detail and demonstrated for the Cu-Mg system down to 0 K using unary descriptions based on segmented regression. The model parameter uncertainties are quantified and propagated to the Gibbs energy functions.

* Corresponding author. E-mail: bjb54@psu.edu,

# 1 Introduction

Thermodynamics is the foundation of science that concerns the state of a system under given external conditions. The CALPHAD method describes the thermodynamic behavior of a system by a semi-empirical parametric model of the Gibbs energy of each phase in the system. Developing models and evaluating model parameters is central to the CALPHAD method, but, with the exception of some recent attempts to study the influence of model parameterization in CALPHAD models computationally, the state of the art user tools [1,2] for fitting CALPHAD model parameters have remained relatively stagnant over the years.

Following the publication of the SGTE description of the unary systems [3], significant effort has been made to develop multicomponent databases based on descriptions of the common constituent subsystems. This has led to the convergence of the CALPHAD community on certain key unary and binary descriptions, which adds considerable inertia to updating the descriptions with the most recent scientific understanding. For example, new unary descriptions are being developed that extend down to 0 K using the Einstein [4] or Debye [5] models, include better descriptions of the high temperature behavior of the heat capacity for solid phases, or that take into account variability in the reported data by outlier detection [6] or automatic weighting schemes. In addition, only recently has the topologically close-packed σ phase been modeled with five sublattices [7–9] corresponding to the five Wyckoff positions in that phase, but nonetheless there are still new assessments published where the σ phase is described with fewer sublattices to stay compatible with previous and future assessments [10,11].

With the rapid increase of available computational data and the advances in physically-based CALPHAD models for Gibbs energies and other physical properties [12], the current methods for optimizing model parameters to data cannot support the growing need to assess and compare different types of models to data and to maintain the existing databases when new data must be incorporated. Furthermore, a key aspect of communicating the efficacy of newly developed models and databases to consumers of CALPHAD databases is to robustly quantify the uncertainty of the model with respect to the underlying data. Therefore, it is necessary to develop a new generation of software tools that can support not only model development, database development and maintenance, but also to quantitatively determine and report the confidence in the optimized models and derived properties to database users.

Here we introduce ESPEI (Extensible Self-optimizing Phase Equilibria Infrastructure) as an open-source, Python-based application for CALPHAD database development and uncertainty quantification. The rest of this paper will develop the theory behind the two fitting modules of ESPEI, single-phase parameter selection and multi-phase Markov Chain Monte Carlo (MCMC) optimization and uncertainty quantification. Finally, ESPEI will be applied to select model parameters and optimize the parameters in the Cu-Mg binary system, propagating the parameter uncertainties to the Gibbs energy functions and the phase equilibria.

## 2 Theory

ESPEI implements two steps of model parameter evaluation: generation and MCMC optimization. The parameter generation step uses [13] experimental and first-principles data describing the derivatives of the Gibbs free energy to parameterize the Gibbs energies of each individual phase within the Compound Energy Formalism (CEF)[14], giving the user a complete thermodynamic database based only on thermochemical data. Experimental thermochemical data for virtually all real alloy systems are too sparse to fully describe the Gibbs energies of the phases and are often unable to access the energetics of metastable configurations defined within the CEF. In practice, the CALPHAD method requires that available experimental data is augmented with thermochemical data from estimates, empirical models [15], machine learning models [16], or first-principles calculations [17], all of which only give approximate energies. Therefore, assessments based on thermochemical data alone cannot reproduce the true Gibbs energies of each phase and, by extension, the phase diagram and other thermodynamic properties. For this reason, the CALPHAD method has modeled the Gibbs energies by fitting to all available thermodynamic data, including both thermochemical and phase equilibrium data, since its inception. The phase equilibrium data require the Gibbs energies of all phases to be defined, so parameters must be optimized iteratively to be self-consistent by the modeler. The MCMC optimization implemented in ESPEI fulfills the role of iterative, self-consistent optimization of Gibbs energy parameters considering all data simultaneously. The following sections will discuss each fitting step in detail.

## 2.1 Parameter selection from thermochemical data

Implementing parameter selection with only thermochemical data allows the Gibbs energy models of each phases to be evaluated directly from the data, therefore models for each phase can be generated independently. The construction of the Gibbs energy within the CEF involves the extrapolation of the low order terms in composition into the interaction space, so all endmembers must be fit first, followed by the excess parameters from low order to high order. Interaction parameters are evaluated with candidate models from zeroth order to fourth order interactions, fitting all the temperature dependent terms in a Redlich-Kister polynomial for the zeroth order interaction before fitting all the Redlich-Kister terms for the first order interaction and so on. Endmember energies and interactions are evaluated in the following sequence, using a hypothetical three sublattice ternary solution phase as a prototype for a phase with v-sublattices and j-components.

1. Fit 1-sublattice binary interactions, e.g. (A, B) : * : *, with * being any component
2. Fit 2-sublattice binary interactions, e.g. (A, B) : (*, *) : *
3. Fit v-sublattice binary interactions
4. Fit 1-sublattice ternary interactions, e.g. (A, B, C) : * : *
5. Fit 2-sublattice ternary interactions, e.g. (A, B, C) : (*, *, *) : *
6. Fit v-sublattice, j-component interactions

A simple model with few parameters is better than a complex model that describes the same data marginally better. Parameter generation in ESPEI aims to achieve a balance of a simple parameterization and goodness of fit of the model to the data. The polynomial for the Gibbs energy as a function of temperature has the form given in Eq. 1. Different parameterizations combinations are considered, such as a parameterization of only $a$, only $a$ and $b$, and so on. The parameters that will be considered are limited by the input thermochemical data and how the derivatives of the Gibbs energy can describe the parameters. In general, candidate models that could not physically describe the underlying Gibbs energy function are not generated. For example, isothermal enthalpy of formation data cannot describe high order terms of the Gibbs energy, and thus they will not be considered when only enthalpy data is used.

$$G = a + bT + cT \ln T + \sum_n d_n T^n \qquad \text{Eq. 1}$$

In this polynomial, all the terms are linear with respect to the model coefficients, so the candidate models can be fit to input thermochemical data by a linear least squares optimization under the assumption that the residuals follow a Gaussian distribution. A tunable hyperparameter for L2 regularization that penalizes the magnitude of the parameters can be adjusted by the user to prevent overfitting, resulting in the objective function given in Eq. 2, where $w$ is the vector of coefficients, $X$ the matrix of polynomial terms, $y$ the data to be fit and $\alpha$ the hyperparameter controlling the mixing between the fit to the data and the magnitude of the parameters.

For each fitting step, the parameters for are the candidate models are evaluated by minimizing Eq. 2, then the residual sum of squares (RSS) between the evaluated parameters and the data are used to score and compare the models within the corrected Akaike information criterion (AICc) [18]. The model with the lowest score is the optimal combination of model fitness and complexity. The AICc is a modified version of the AIC that avoids overparameterization when the data is sparse, which is often the case for thermochemical data. The AICc score is calculated by Eq. 3, where $k$ is the number of parameters, $n$ the number of samples, and RSS is the residual sum of squares between the model predictions and the data. In this way, models for each phase with a reasonable fit/complexity tradeoff can be built up one parameter at a time.

$$O = (Xw - y)^2 + \alpha w^2 \qquad \text{Eq. 2}$$

$$\text{AICc} = \text{AIC} + \frac{2k^2 + 2k}{n - k - 1} = 2k - n \ln\frac{\text{RSS}}{n} + \frac{2k^2 + 2k}{n - k - 1} \qquad \text{Eq. 3}$$

2.2 Parameter optimization to phase equilibria data

The CEF uses sublattices to describe the chemical behavior of occupancy and substation of elements on sublattices, typically representing the Wyckoff positions in crystalline materials. The endmembers that the CEF introduces into the Gibbs energy model are often not stable or cannot be characterized completely experimentally due to the high dimensionality of the sublattice composition space that is not directly accessible by experiments. Measurements of the Gibbs energy derivatives in CEF models cannot be obtained completely from experiments in the general case, requiring the use of thermodynamic data indirectly related to the Gibbs energies of the phases to refine the parameters generated from thermochemical data to construct a self-consistent set of parameters that reproduce the known data for a given system. In many practical binary and ternary systems of interest, a typical CALPHAD description might have on the order of tens to hundreds of parameters to assess. Current optimization methods require modelers to restrict the degrees of freedom to only a handful of simultaneous parameters and optimize the restricted sub-problems iteratively.

Bayesian optimization is a method of optimizing model parameters by Bayes Theorem [19], obtaining the posterior probability, $p(\theta|D)$, of the model parameters, $\theta$, conditioned on the data, $D$, in Eq. 4. The prior, $p(\theta)$, contains the domain knowledge of the modeler in the probability distribution of each parameter. The likelihood, $p(D|\theta)$, is the probability that the data is described by a set of parameters, and the evidence, $p(D)$, is the probability of the data marginalized over the possible parameters. In CALPHAD databases, evaluating a closed form solution for these quantities over all model parameters is not computationally achievable, so the posterior probability is determined numerically using MCMC.

$$p(\theta|D) = \frac{p(\theta)p(D|\theta)}{p(D)} \qquad \text{Eq. 4}$$

ESPEI uses MCMC to perform a Bayesian optimization of all model parameters simultaneously. In principle, MCMC optimization in ESPEI can be performed for arbitrarily sized multicomponent, multiphase systems with any number of degrees of freedom, however this is not computationally feasible in practice and it is expected that the extrapolation of binary and ternary CALPHAD parameters into multicomponent space precludes the need to model all the degrees of freedom in a multicomponent system simultaneously. However, a challenge in applying MCMC to optimize parameters in CALPHAD databases is that most MCMC samplers assume that model parameters are uncorrelated to efficiently explore parameter space [19]. The parameters in CALPHAD models for each phase are correlated to each other because increasing the value one parameter and decreasing another can give the same Gibbs energy for any given set of conditions. This challenge is addressed in ESPEI by using an ensemble sampler, as introduced by Goodman and Weare [20]. Ensemble samplers use an ensemble of Markov chains to form the proposal distribution for the parameters. This allows the proposals to be invariant under affine transformations, solving the problems of scaling proposal length and differing parameter magnitudes in multidimensional parameter space simultaneously. Proposed parameters are accepted or rejected based on the Metropolis criteria. The probability of accepting a step is given in Eq. 5, where $p_{\text{current}} = p(\theta_i)p(D|\theta_i)$ is the probability for the current set of parameters and $p_{\text{proposed}} = p(\theta_{i+1})p(D|\theta_{i+1})$ is the probability for the newly proposed parameters. Thus, the Metropolis criteria accepts all proposed parameters that increase the probability, i.e. $\frac{p_{\text{proposed}}}{p_{\text{current}}} > 1$ so $p_{\text{accept}} = 1$, but there's also a $\frac{p_{\text{proposed}}}{p_{\text{current}}}$ chance of accepting parameters that decrease the probability accepts proposals that decrease the posterior probability. The ability to accept parameters that decrease the probability leads to the convergence of the probability to the true posterior distribution.

$$p_{\text{accept}} = \min\left(\frac{p_{\text{proposed}}}{p_{\text{current}}}, 1\right) \qquad \textbf{Eq. 5}$$

Adding parameters that decrease the posterior probability to the Markov chain systematically constructs the posterior distribution from the point estimates of the probability density. ESPEI uses an ensemble sampler algorithm implemented in the emcee package [21] that implements parallelizable ensemble samplers. ESPEI provides emcee with an initial ensemble of chains as Gaussian distributions centered on the parameters generated by single phase fitting and defines a probability function that calculates point posterior log-probabilities from the prior and likelihood. Log-probabilities are often used in MCMC software to prevent floating point errors when multiplying probabilities. It should be noted that ESPEI can fit parameters for any type of model in a thermodynamic database if the individual model parameters can be specified using the FUNCTION command in the thermodynamic database (TDB) format.

Prior distributions for the parameters are the main way that modelers input domain knowledge into ESPEI's MCMC optimization. Each parameter has a probability density function associated with it that, for a given set of parameters, can be used to evaluate $p(\theta)$. Often the same type of prior distribution is used for each parameter, but this is not a requirement. Any function can be used in ESPEI, though there are three prior distribution functions that are easy to use in ESPEI: a uniform, normal, or triangular prior. For convenience, ESPEI provides a method for describing the hyperparameters for these distributions relative to the parameters in the initial database. Uniform priors are uninformative, giving constant probability between upper and lower parameter bounds and are useful when there is low confidence in the initial database parameters. Normal priors follow a Gaussian distribution, usually centered on the parameter. Normal priors are useful for informing a centrality of each parameter in the prior, but not explicitly limiting the parameter bounds. Triangular priors are particularly useful for modeling in ESPEI, as they allow the modeler to inform the centrality of the parameters, e.g. from parameter selection, but also to enforce bounds where the prior probability is zero outside of those bounds, which can be used to ensure that parameters stay the same sign or within the same order of magnitude, for example. The hyperparameters to generate the prior distributions are specified in ESPEI input files. A detailed example and further explanation are included in the ESPEI software documentation [22].

Three main types of data are considered by the likelihood function defined in ESPEI: single phase thermochemical data of the temperature derivatives of the Gibbs energy, activity data, and multi-phase equilibria data. To ensure consistent weighting of error from different data types, the likelihood for each type of data is normalized by the standard deviation of the error for that data type. For all data types, the error is assumed to follow a normal distribution. ESPEI provides default values for the standard deviations of each type of data, however users can modify the values by adding a weight for each type of data or for each individual dataset. The thermodynamic quantities for each type of data are determined in ESPEI by using pycalphad [23] as the thermodynamic calculation engine.

The error due to thermochemical data is calculated by comparing the difference between the predicted values and the measured or calculated values. Since these values are calculated or measured directly, the standard deviations of the error are relatively well defined. The default standard deviations for enthalpy, entropy, and heat capacity data are chosen to be 500 J/mol, 0.2 J/K-mol, and 0.2 J/K-mol, respectively. To fit activity data, the measured activity is converted to chemical potentials as suggested by Lukas *et al.* [24]. The error is calculated by the difference between calculated and measured chemical potentials. Although activities are not measured directly, the experimental errors are mathematically related to the activity and the experimental errors reported in the literature can be propagated to the chemical potentials, ESPEI chooses a standard deviation in the chemical potential due to measured activity data of 500 J/mol.

Thermodynamic assessments are judged, to the first order, by the agreement with the measured phase equilibria on the phase diagram, so any complete thermodynamic assessment program must consider optimizing to phase equilibria data. The phase equilibria depend on solving for the global minimum energy for many phases and arbitrary models of the Gibbs energy, so a solution mapping the model parameters to the calculated equilibria is not well defined. Several methods exist for determining the error for phase equilibria. The direct approach is to calculate the phase equilibria and determine the error in the phase boundary due to a different in temperature or composition. However, a drawback of the direct approach is that the phase equilibria of interest are not always stable and a near-optimal solution must already be known to properly calculate the error. Since MCMC optimization considers many degrees of freedom and the initial database may not be close to a globally optimal solution, a more generic approach is warranted. The method implemented by ESPEI is similar to the rough search method implemented in the PanOptimizer software [2]. Given a set of measured phase constitutions in equilibrium (e.g. from EPMA) and a set of candidate parameters for a model the multiphase error can be calculated as follows (illustrated schematically in Figure 1):

1. At each measured tieline vertex, perform equilibrium calculations with all phases active. Construct the target equilibrium hyperplane as the arithmetic mean of the calculated chemical potentials, shown as a solid line in Figure 1a.
2. Compute the single-phase Gibbs energy of each tie vertex, $G^t$, with only the desired phase active. These effectively form the current hyperplane of the given parameters, shown in Figure 1b.
3. The residual is the driving force between the target and current hyperplane, calculated as $G^t - \sum_i \bar{\mu}_i x_i$, where $\bar{\mu}_i$ is the chemical potential of the target hyperplane calculated in step 1.

This approach is advantageous because it generalizes well to multicomponent systems and can be used even when the measured phases in equilibrium are not in the calculated stable equilibrium, however there are some special cases that must be considered. It is sometimes the case, especially in binary systems, that the equilibrium phase boundary compositions are unknown for some tie vertices, such as solidus and liquidus boundaries determined by heating and cooling curves. These unknown compositions are excluded from the determination of the target hyperplane. Stoichiometric compounds must be excluded as tie vertices when determining the target hyperplane because the equilibrium chemical potential is discontinuous at the composition of the compound. Using the prior and likelihood functions ESPEI calculates the probabilities for a proposed set of parameters and through MCMC, the posterior probability distributions for each parameter that maximize the probability that the models describe the experimental and calculated data.

## 3   Cu-Mg system

The Cu-Mg system exhibits features found in most complex CALPHAD databases including terminal phases with solubility, a stoichiometric compound, and an intermetallic compound with solubility. This binary system is a technologically important system for bulk metallic glass (BMG) systems [25] that are produced using molds that are actively cooled, for example with liquid $N_2$ and for Al-based alloys [26]. CALPHAD models that extend down to low temperatures can be used to describe the energetics for phase precipitation in potential BMG alloys. Because ESPEI can rapidly select and fit different CALPHAD models, it is ideal for comparing different models and unary reference states. Here ESPEI will be used to develop a description for the Cu-Mg system that extends down to 0 K using segmented regression models for the Cu and Mg unary descriptions [5]. The equilibrium Cu-Mg system has four stable solid phases: fcc and hcp solutions, an intermetallic compound, $CuMg_2$, and a C15-type $Cu_2Mg$ Laves phase with considerable solubility. The Cu-Mg system has previously been modeled by several times [27,28], most recently by Zhou *et al.* [29][30]. Interested readers are directed to those publications for a thorough review of the existing data.

Recently, first-principles calculations for the terminal solution phases in the Cu-Mg system were calculated by Gao *et al.* [31]. These data are also considered in our parameter selection and optimization. The latest programming interface, command line usage, and additional documentation including code examples, a changelog of ESPEI versions and links to the development repository can be found at https://espei.org [22]. The documentation and user manual for ESPEI version 0.6.2 is included in the supplemental material.

3.1 Unary reference data

The segmented regression (SR) model [5] is one of the alternative physically-based formulations to the recently used SGTE description of reference data for pure elements. The SR model has been developed for the description of temperature dependence in the heat capacities down to 0K considering relevant physical effects and it consists of three terms:

$$C_P(T,\boldsymbol{\theta}) = \frac{9N_A k_B}{(\theta_D/T)^3} \int_0^{\theta_D/T} \frac{x^4 e^{-x}}{(1-e^{-x})^2} dx + \begin{cases} \beta_1 T, & T < \alpha - \gamma \\ \beta_1 T + \beta_2 \frac{(T-\alpha+\gamma)^2}{4\gamma}, & \alpha - \gamma \leq T \leq \alpha + \gamma \\ \beta_1 T + \beta_2 (T-\alpha), & \alpha + \gamma < T \end{cases} + C_P^{magn},$$

**Eq. 6**

where $\boldsymbol{\theta} = (\theta_D, \beta_1, \beta_2, \alpha, \gamma)$ is the vector of unknown model parameters to be estimated from experimental and computational data. Here, the main contribution to the heat capacity due to phonon vibration is represented by the well-known Debye model. The second term in Eq. 6 is a so-called bent-cable model. This term is introduced for the decomposition of linearly-dependent physical effects at low and high temperatures. The last term describes the magnetic contributions to the heat capacity. This term is not considered and equal to zero for non-magnetic elements.

Since the analytical formulation of the Debye model cannot be implemented directly in the TDB format, a non-series approximation of the Debye model base on the weighted linear combination of Einstein functions for heat capacity has been developed to overcome this issue and it has been successfully applied for the thermodynamic re-evaluation of the Cr-Nb system [32] and agrees well with available experimental data over the entire temperature range. The Gibbs energy function for reference data is derived based on the well-known thermodynamic relationship between heat capacity and Gibbs energy,

$$G(T) = \int C_P(T) - T \int \frac{C_P(T)}{T} dT \qquad \text{Eq. 7}$$

The heat capacities fit for Cu and Mg, presented in Figure 2, and the Gibbs energies, presented in Figure 3, show good agreement with experimental data from 0K up to the melting point. At the time of writing, ESPEI does not support fitting unary descriptions of pure elements and by default does not adjust the unary model parameters in the MCMC optimization. Therefore, the uncertainty in different unary descriptions cannot yet be propagated into the binary and higher order systems that ESPEI assesses.

3.2 ESPEI optimization

Parameters were generated for the Cu-Mg system based on the measured and calculated single phase thermochemical data. Experimental enthalpy of formation for the CuMg$_2$ [33,34] and Laves C15 [33,34] phases and experimental enthalpy mixing for the liquid phase [35] were used. First-principles calculations for the mixing energies for the FCC_A1 and HCP_A3 phases were calculated by Shin [36] and Gao *et al.* [37], while the endmembers of the Laves phase and CuMg$_2$ phase were calculated by Zhou *et al.* [29]. In total, 15 degrees of freedom were generated for the endmembers and binary interactions for the five phases. The resulting phase diagram is shown in Figure 4. This phase diagram, produced only from the thermochemical data, shows the qualitative behavior of the measured phase equilibria, even though those data were not included in the optimization. The parameters that produced this phase diagram were used as input for the MCMC optimization step with triangular priors that spanned from $\pm 0.5\theta$. All the available thermochemical, activity and phase equilibrium data were considered, and all parameters fit to the data simultaneously. Figure 5 shows the phase diagram after 500 MCMC iterations and shows excellent agreement with the phase equilibria data.

3.3    Uncertainty quantification

CALPHAD-type calculations could consider parameter uncertainty and propagated uncertainty. Parameter uncertainty concerns the distribution of each parameter in a model, while propagated uncertainty is the parameter uncertainty that has been carried through to the uncertainty in different thermodynamic properties that may be calculated. Individual parameter uncertainties are usually not of interest in isolation because the parameters and uncertainties usually are marginalized across several CALPHAD parameters and models as contributions to the Gibbs energy. Parameter uncertainty is evaluated within the MCMC optimization step of ESPEI by quantifying the distribution of the parameter values that make up each Markov chain. To correctly quantify uncertainty, each chain must be converged. MCMC convergence cannot be proved for arbitrary simulations, however several criteria exist that are indicators of convergence. The application of these indicators is the same in ESPEI as any MCMC simulation and are discussed in detail elsewhere [19].

The liquid parameters from the Markov chain constructed in the MCMC optimization are shown in the corner plot in Figure 6a. The histograms for each parameter are plotted along the diagonal of the corner plot and the covariance between each pair of parameters is plotted below the diagonal. The histograms of each parameter are not required to follow any distribution because the probability is the product of the likelihood and the prior, which can have varying shape in parameter space. For parameters in CALPHAD models, it is expected that some parameters will exhibit covariance, particularly for parameters in the same phase and between phases in equilibrium. Too many parameters that are correlated are an indication that fewer parameters are required to describe the Gibbs energies that the data require. Uncertainty from the MCMC chains and probabilities were propagated to the Gibbs energy of mixing of the liquid phase in Figure 6b. To propagate the uncertainty, 95% of the parameter values with the highest probability density were used to calculate the Gibbs energies and phase diagrams, forming the 95% highest density intervals (HDI) for each type of data. The mean Gibbs energy is plotted as the line, and the 95% HDI is plotted as the shaded region in Figure 6b. The uncertainty in the Gibbs energies of the liquid region are smaller where liquid is in stable two phase equilibrium on the phase diagram and where the different sets of experiments agree. Near the congruent melting of the Laves phase, there is some disagreement in the data, which leads to a region of relatively larger uncertainty. The maximum uncertainty is on the order of 1 kJ/mol, which is consistent with what would be expected from experiments. Since the unary descriptions are fixed in this optimization, there is no uncertainty bound at either pure Cu or pure Mg will always be 0. Fitting unary parameters in ESPEI would enable the uncertainty quantification. More details on the statistical approach of propagating parameter uncertainty, especially to phase diagrams, will be published in a separate article [38].

# 4 Conclusion

ESPEI allows thermodynamic databases to be rapidly generated using thermochemical data, then refined considering all thermodynamic data for a system simultaneously with little user interaction. The parameter selection and MCMC optimization steps in ESPEI make it the first CALPHAD-based software that can generate model parameters, adjust the parameters, and quantify uncertainty for arbitrarily sized multicomponent systems. The two-step optimization approach allows existing and newly created databases to be updated when new experimental or calculated data becomes available. The Cu-Mg system was assessed using ESPEI and a new method for depicting phase diagram uncertainty was presented. ESPEI is developed in the open with the intent that it become adopted as a platform for the development and assessment of Gibbs energy models and other property models.

# 5 Acknowledgements

The authors thank Jiong Wang, ShunLi Shang, and Yi Wang for their help in testing ESPEI. B.B. was supported by a NSF National Research Trainee Fellowship grant DGE-1449785. This work was also supported by a NASA Space Technology Research Fellowship grant number 80NSSC18K1168 and used the NSF Extreme Science and Engineering Discovery Environment (XSEDE), which is supported by NSF grant number ACI-1548562. R.O. acknowledges financial support from NASA's Science Mission Directorate and Space Technology Mission Directorate through the Game Changing Development program under Prime Contract #80NM0018D0004, and the Space Technology Office at the Jet Propulsion Laboratory, California Institute of Technology. The financial support from the Collaborative Research Center "Superalloys Single

Crystal" (TR-103 project C6) of the German Research Foundation (DFG) and the Sino-German Cooperation Group is acknowledged.

# 6 References

[1] J.O. Andersson, T. Helander, L. Höglund, P. Shi, B. Sundman, Thermo-Calc & DICTRA, computational tools for materials science, Calphad. 26 (2002) 273–312. doi:10.1016/S0364-5916(02)00037-8.

[2] W. Cao, S.L. Chen, F. Zhang, K. Wu, Y. Yang, Y.A. Chang, R. Schmid-Fetzer, W.A. Oates, PANDAT software with PanEngine, PanOptimizer and PanPrecipitation for multi-component phase diagram calculation and materials property simulation, Calphad Comput. Coupling Phase Diagrams Thermochem. 33 (2009) 328–342. doi:10.1016/j.calphad.2008.08.004.

[3] A.T. Dinsdale, SGTE data for pure elements, Calphad. 15 (1991) 317–425. doi:10.1016/0364-5916(91)90030-N.

[4] S. Bigdeli, H. Mao, M. Selleby, On the third-generation Calphad databases: An updated description of Mn, Phys. Status Solidi Basic Res. 252 (2015) 2199–2208. doi:10.1002/pssb.201552203.

[5] I. Roslyakova, B. Sundman, H. Dette, L. Zhang, I. Steinbach, Modeling of Gibbs energies of pure elements down to 0 K using segmented regression, Calphad Comput. Coupling Phase Diagrams Thermochem. 55 (2016) 165–180. doi:10.1016/j.calphad.2016.09.001.

[6] N.H. Paulson, E. Jennings, M. Stan, Bayesian strategies for uncertainty quantification of the thermodynamic properties of materials, (2018). http://arxiv.org/abs/1809.07365.

[7] Z. Li, H. Mao, P.A. Korzhavyi, M. Selleby, Thermodynamic re-assessment of the Co-Cr

system supported by first-principles calculations, Calphad Comput. Coupling Phase Diagrams Thermochem. 52 (2016) 1–7. doi:10.1016/j.calphad.2015.10.013.

[8] N. Dupin, U.R. Kattner, B. Sundman, M. Palumbo, S.G. Fries, Implementation of an Effective Bond Energy Formalism in the Multicomponent Calphad Approach, J. Res. Natl. Inst. Stand. Technol. 123 (2018) 123020. doi:10.6028/jres.123.020.

[9] R. Mathieu, N. Dupin, J.-C. Crivello, K. Yaqoob, A. Breidi, J.-M. Fiorani, N. David, J.-M. Joubert, CALPHAD description of the Mo–Re system focused on the sigma phase modeling, Calphad. 43 (2013) 18–31. doi:10.1016/j.calphad.2013.08.002.

[10] W.M. Choi, Y.H. Jo, D.G. Kim, S.S. Sohn, S. Lee, B.J. Lee, A Thermodynamic Modelling of the Stability of Sigma Phase in the Cr-Fe-Ni-V High-Entropy Alloy System, J. Phase Equilibria Diffus. 39 (2018) 694–701. doi:10.1007/s11669-018-0672-x.

[11] J.-M. Joubert, J.-C. Crivello, Non-Stoichiometry and Calphad Modeling of Frank-Kasper Phases, Appl. Sci. 2 (2012) 669–681. doi:10.3390/app2030669.

[12] C. Marker, S.L. Shang, J.C. Zhao, Z.K. Liu, Elastic knowledge base of bcc Ti alloys from first-principles calculations and CALPHAD-based modeling, Comput. Mater. Sci. 140 (2017) 121–139. doi:10.1016/J.COMMATSCI.2017.08.037.

[13] Z.K. Liu, First-principles calculations and CALPHAD modeling of thermodynamics, J. Phase Equilibria Diffus. 30 (2009) 517–534. doi:10.1007/s11669-009-9570-6.

[14] M. Hillert, The compound energy formalism, J. Alloys Compd. 320 (2001) 161–176. doi:10.1016/S0925-8388(00)01481-X.

[15] F.R. De Boer, W.C.M. Mattens, R. Boom, A.R. Miedema, A.K. Niessen, Cohesion in metals, North-Holland, 1988.

[16] G. Hautier, C.C. Fischer, A. Jain, T. Mueller, G. Ceder, Finding natures missing ternary

oxide compounds using machine learning and density functional theory, Chem. Mater. 22 (2010) 3762–3767. doi:10.1021/cm100795d.

[17] F. Liu, S.J. Song, F. Sommer, E.J. Mittemeijer, Evaluation of the maximum transformation rate for analyzing solid-state phase transformation kinetics, Acta Mater. 57 (2009) 6176–6190. doi:10.1016/j.actamat.2009.08.046.

[18] J.E. Cavanaugh, Unifying the derivations for the Akaike and corrected Akaike information criteria, Stat. Probab. Lett. 33 (1997) 201–208. doi:10.1016/S0167-7152(96)00128-9.

[19] A. Gelman, H.S. Stern, J.B. Carlin, D.B. Dunson, A. Vehtari, D.B. Rubin, Bayesian data analysis, Chapman and Hall/CRC, 2013.

[20] J. Goodman, J. Weare, Ensemble samplers with affine invariance, Commun. Appl. Math. Comput. Sci. 5 (2010) 65–80. doi:10.2140/camcos.2010.5.65.

[21] D. Foreman-Mackey, D.W. Hogg, D. Lang, J. Goodman, emcee : The MCMC Hammer, Publ. Astron. Soc. Pacific. 125 (2013) 306–312. doi:10.1086/670067.

[22] B. Bocklund, ESPEI Software Documentation, (2019). https://espei.org.

[23] R. Otis, Z.-K. Liu, pycalphad: CALPHAD-based Computational Thermodynamics in Python, J. Open Res. Softw. 5 (2017) 1. doi:10.5334/jors.140.

[24] H. Lukas, S.G. Fries, B. Sundman, Computational Thermodynamics The Calphad Method, Cambridge University Press, 2007. doi:10.1017/CBO9780511804137.

[25] N.P. Bailey, J. Schiøtz, K.W. Jacobsen, Simulation of Cu-Mg metallic glass: Thermodynamics and structure, Phys. Rev. B - Condens. Matter Mater. Phys. 69 (2004) 1–11. doi:10.1103/PhysRevB.69.144205.

[26] T. Buhler, S.G. Fries, P.J. Spencer, H.L. Lukas, A thermodynamic assessment of the Al-Cu-Mg ternary system, J. Phase Equilibria. 19 (1998) 317–329.

doi:10.1361/105497198770342058.

[27] C.A. Coughanowr, I. Ansara, R. Luoma, M. Hamalainen, H.L. Lukas, Assessment of the Cu-Mg system, Zeitschrift F{ü}r Met. 82 (1991) 574–581.

[28] Y. Zuo, Y.A. Chang, Thermodynamic calculation of the Mg-Cu phase diagram, Zeitschrift Für Met. 84 (1993) 662–667.

[29] S. Zhou, Y. Wang, F.G. Shi, F. Sommer, L.-Q. Chen, Z.-K. Liu, R.E. Napolitano, Modeling of Thermodynamic Properties and Phase Equilibria for the Cu-Mg Binary System, J. Phase Equilibria Diffus. 28 (2007) 158–166. doi:10.1007/s11669-007-9022-0.

[30] S. Zhou, Y. Wang, F.G. Shi, F. Sommer, L.-Q. Chen, Z.-K. Liu, R.E. Napolitano, Modeling of thermodynamic properties and phase equilibria for the Cu-Mg binary system, J. PHASE EQUILIBRIA Diffus. 28 (2007) 158–166. doi:10.1007/s11669-007-9022-0.

[31] Q. Gao, J. Wang, S. Shang, S. Liu, Y. Du, Z.-K. Liu, First-principles calculations of finite-temperature thermodynamic properties of binary solid solutions in the Al–Cu–Mg system, Calphad. 47 (2014) 196–210. doi:10.1016/j.calphad.2014.10.004.

[32] Y. Jiang, S. Zomorodpoosh, I. Roslyakova, L. Zhang, Thermodynamic re-assessment of binary Cr-Nb system down to 0 K, Calphad Comput. Coupling Phase Diagrams Thermochem. 62 (2018) 109–118. doi:10.1016/j.calphad.2018.06.001.

[33] H. Feufel, F. Sommer, Thermodynamic investigations of binary liquid and solid Cu-Mg and Mg-Ni alloys and ternary liquid Cu-Mg-Ni alloys, J. Alloys Compd. 224 (1995) 42–54. doi:10.1016/0925-8388(95)01526-4.

[34] R. King, O. Kleppa, A thermochemical study of some selected laves phases, Acta Metall. 12 (1964) 87–97. doi:10.1016/0001-6160(64)90056-2.

[35] G.I. Batalin, V.S. Sudavtsova, M.V. Mikhailovskaya, Thermodynamic Properties of

Liquid Alloys of the Cu–Mg Systems, Izv. Vyss. Ucheb. Zaved., Tsvetn. Met. 2 (1987) 29–31.

[36] D. Shin, Thermodynamic properties of solid solutions from special quasirandom structures and CALPHAD modeling: Application to aluminum-copper-magnesium-silicon and hafnium-silicon-oxygen, The Pennsylvania State University, 2007.

[37] Q.N. Gao, J. Wang, S.L. Shang, S.H. Liu, Y. Du, Z.K. Liu, First-principles calculations of finite-temperature thermodynamic properties of binary solid solutions in the Al-Cu-Mg system, Calphad. 47 (2014) 196–210. doi:10.1016/j.calphad.2014.10.004.

[38] N.H. Paulson, B.J. Bockund, R.A. Otis, Z.-K. Liu, M. Stan, Representation of Quantified Uncertainty in CALPHAD for Materials Design, Prep. (n.d.).

## 7 Figures

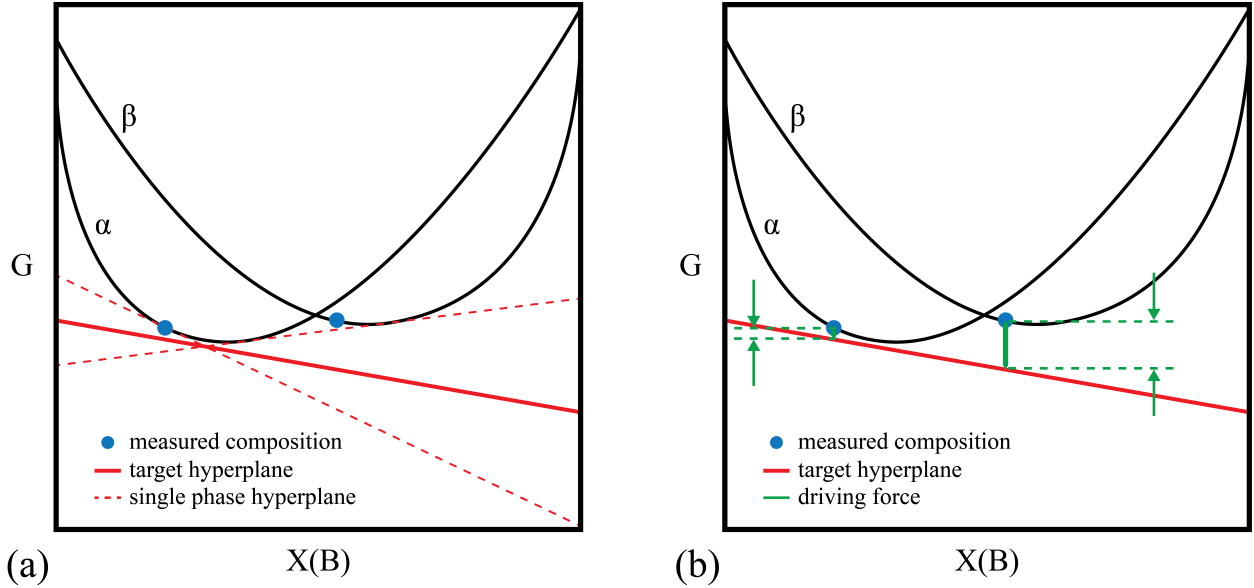

**Figure 1.** Schematic procedure for calculating the error in phase equilibrium data. *α* and *β* are Gibbs energy curves for hypothetical phases. The blue points represent measured phase constitutions projected onto the Gibbs energy curves. The solid red line represents the target hyperplane as the arithmetic mean of the hyperplanes for each tieline vertex (red dashed lines in (a)). The driving force error (b) is the difference between the target hyperplane and the current hyperplane at each vertex.

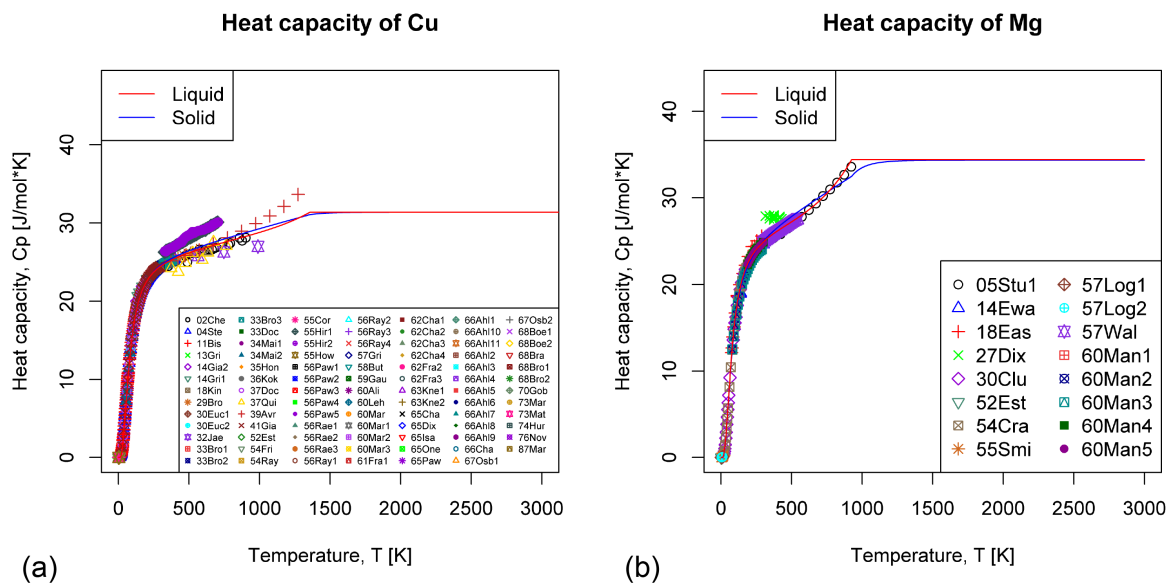

**Figure 2.** Heat capacity of the optimized segmented regression unary models for Cu (a) and Mg (b) compared to the data available in the NIST repository.

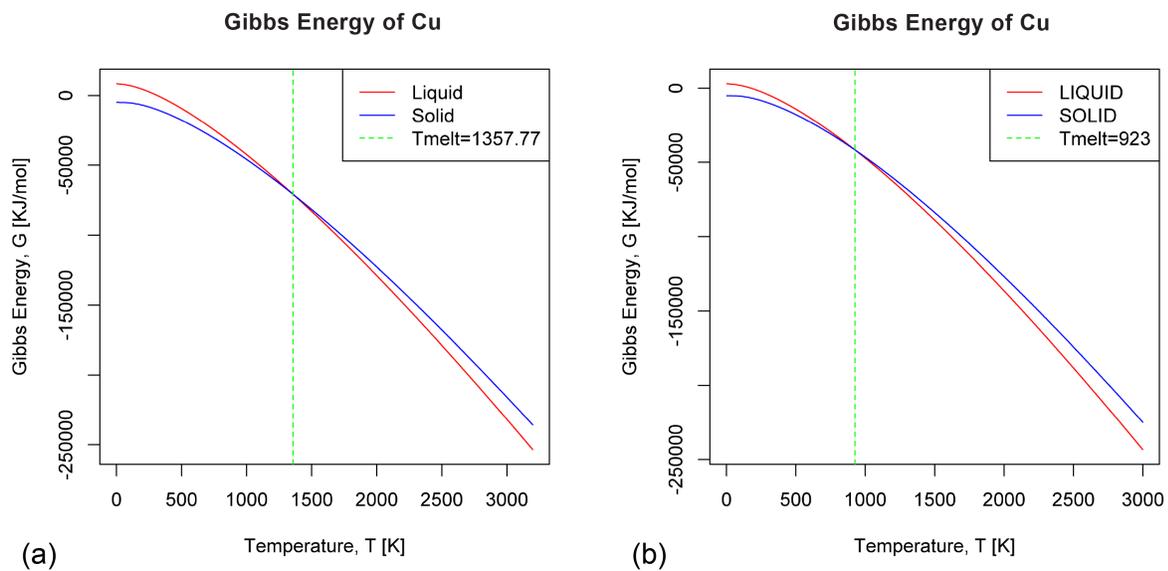

**Figure 3.** Gibbs energy of the optimized segmented regression unary models for Cu (a) and Mg (b). The vertical lines compare the melting points from the SGTE unary data and the melting points as assessed here.

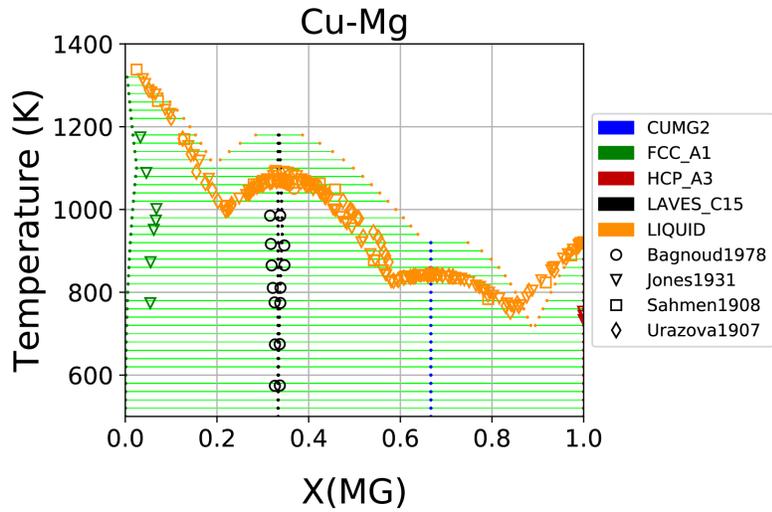

**Figure 4.** Cu-Mg phase diagram as fit by parameter selection and compared to experimental phase equilibrium data.

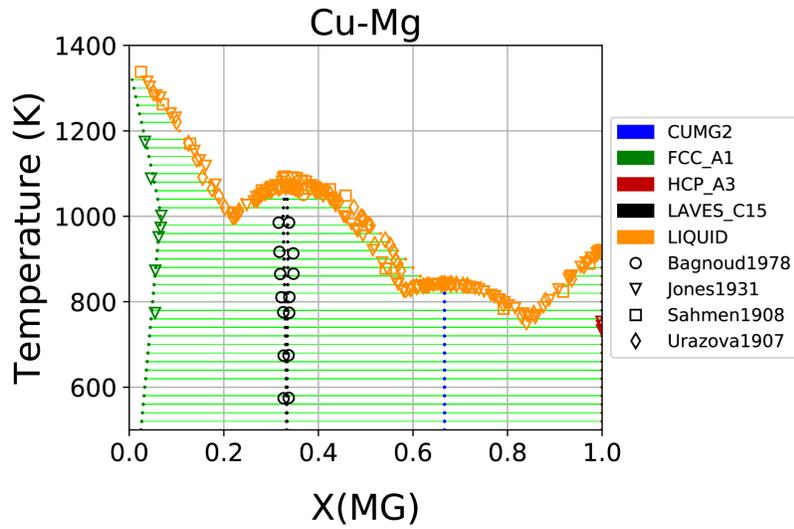

**Figure 5.** Cu-Mg phase diagram compared to experimental phase equilibrium data after MCMC fitting with all data considered simultaneously.

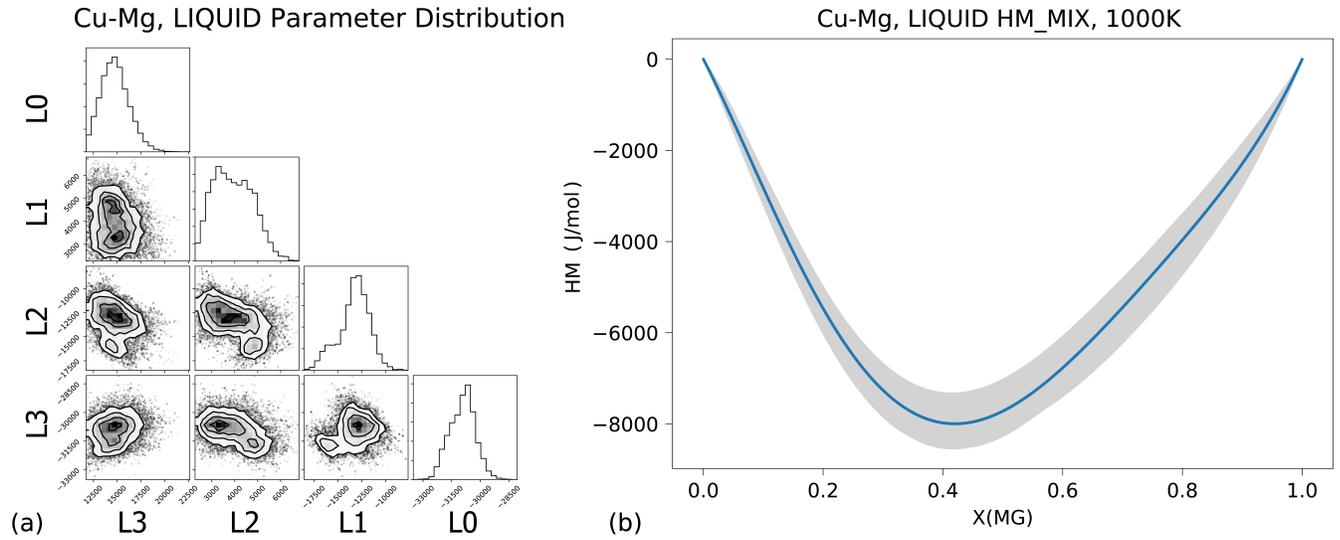

**Figure 6.** Corner plot (a) of the parameters in the FCC phase. The diagonal images show the histogram of each parameter in the Markov chain and the off-diagonal images show the covariance between two parameters. Parameter uncertainty propagated to (b) the Gibbs energy of mixing for the liquid phase. In the blue line in (b) is the mean and the grey area is the 95% high density interval.